\documentclass[article]{elsarticle}

\journal{Science of the Total Environment}

\begin{document}

\begin{frontmatter}

\title{Discussion on the ``Dynamic model to conceptualize multiple environmental
pathways to the epidemic of Chronic Kidney Disease of unknown etiology (CKDu)''}

\author{M. W. C. Dharma-wardana,
National Research Council of Canada
1200, Montreal Rd, Ottawa, Canada K1K 3G5}  



\address[mysecondaryaddress]{Dept. of Physics, Universiti\'{e} de Montreal, Montreal, PQ, Canada.}

\begin{abstract}Jayasinghe et al. [Science of the Total Environment, 705 (2020) 135766] propose a `dynamical' model of Chronic Kidney Disease of Unknown etiology (CKDu) wherein CKDu arises as an emergent property of a complex system where they claim that  weak multiple environmental factors contribute. They criticize the usual approaches as being ``reductionist''. We use their model as a basis of a discussion on the possibility of  treating CKDu as an emergent property resulting from the interaction of multiple weak  environmental factors  with the organism. The model does not reveal anything beyond what is already known from simple considerations of well-known feed-back loops, but has the merit of re-stating those issues in a different format. If a proper weighting of the possible environmental factors is included, most proposed environmental factors drop out and what Jayasinghe et al. call the ``reductionist''  approach naturally arises due to the weight of evidence.  The theory that the consumption of water containing fluoride and magnesium ions as found in water from regolith aquifers drawn via house-hold wells is found to clearly hold within this model when proper weighting is included. However, we show by a counter-example that such models can be easily misused, leading  to completely misleading conclusions. A response formalism useful in the theory of complex systems and emergent modes is presented in the context of the current problem. In addition to there being  a lack of adequate data to fully implement such a theory, it is seen that such elaborations are  unnecessary and misleading in the present context.

\end{abstract}

\begin{keyword}
\texttt{Chronic kidney disease\sep CKDu\sep Systems approach \sep emergent properties
\sep fluoride \sep ground water}
\end{keyword}

\end{frontmatter}


\section{Introduction}Jayasinghe et al.~\cite{Jayasinghe2020} begin
 their short communication  with a review of the
 literature on chronic kidney disease of unknown etiology (CKDu) that affects
tropical pastoral  communities in many parts of the world stretching from meso-American
countries like Ecuador, 
 to Andra Pradesh in India, El-Minia in Egypt  and  to villages scattered mainly
 in the north central province
 of Sri Lanka~\cite{chatterj16}. While Ecuador is the most affected by CKDu, Sri Lanka~\citep{Ranasinghe19}
has led the way in launching a concerted program of research into CKDu since 2008.

 When CKDu  caught the serious attention
of researchers (in early work like  ~\cite{Tabesekera96,Bandara08}), it was natural to 
begin to look at well-known nephrotoxins
 like heavy metals  (e.g., Cd, Pb, As  etc.), pesticides, chlorohydrocarbon residues,
 or even dehydration of workers under field conditions as possibilities. The latter
possibility has now been more or less eliminated in the context of Sri Lanka while the claimed presence
of heavy-metal toxins in  fish~\cite{Bandara08} etc., could not be confirmed by subsequent work.
 Other possible causes~\cite{WHO2}, e.g., snake bite, algal toxins in drinking water, and synergistic
 action of geologically
 occurring fluoride ions together with magnesium ions found in hard water~\cite{DW2017},
 fluoride acting synergistically with Na$^+$ ions~\cite{ChandrajithDissa11}
 or Al$^{3+}$ ions ~\cite{IllepAl},
glyphosate residues acting synergistically  with arsenic and Ca$^{2+}$ ions~\cite{TaskForceGly},
have been proposed. 

Important controlling factors  for the leaching of fluoride from  groundwater aquifers
 are the presence of magnesium and bicarbonates, evapo-transpiration, and long contact times of water
 with the aquifer. The proposal that
 fluoride ions act synergistically with magnesium ions
is based on experimental evidence for characteristic renal-tubular damage in the kidney observed
in laboratory mice~\cite{SynergyBandara2017}, as well as on the positions of fluoride and magnesium ions in
the Hofmeister series~\cite{DW2015}.   Supporting evidence based on Gibbs Free-energy
calculations showing synergies between fluoride and magnesium ions is also available~\cite{DW2017}.
 The fluoride
theory is also supported by the observation of skeletal fluorosis in many residents of endemic
areas. The bicarbonate in hard water found in disease-endemic
areas helps to leach fluorides
from the regolith aquifers that supply the well water used by residents in endemic regions. These wells
are not connected to the agricultural water table linked to water reservoirs (``tanks'') and irrigation
works, as shown by isotopic tracer  studies~\cite{Manthrithilake2016}. 
 Independent studies have shown a correlation of CKDu with groundwater usage~\citep{Kafle19}. 
No CKDu is found in dry-zone farming
areas like the Jaffna peninsula where the drinking water is located  in limestone aquifers,
 while all other factors
discussed by Jayassinghe et al. hold. Similarly,  villages in the region that use water from
 natural springs are also free of CKDu~\cite{DW2015}. Rats fed with dug-well water from endemic areas          
developed renal damage~\citep{Thammiti17}.
Fernanado et al.~\citep{Fernando19} have found
higher levels of fluoride  in serum and urine in endemic CKDu patients as compared with endemic control groups
who live in high fluoride and hard water areas.  
Thus the fluoride-magnesium hypothesis  has a strong  experimental and theoretical basis. 
 
Chronic kidney disease in Andra Pradesh is believed to be similar to Sri Lankan CKDu in its clinical
characteristics. Hence its etiology may also be similar to 
Sri-Lankan Regolith Aquifer-Sourced  Nephropathy (RASN), owing to similar geochemical characteristics
in the two regions~\cite{SubbaRao08}. 

The common pre-conception that agrochemical residues and industrial products in the food chain
(air, water, soil, food) may cause chronic diseases
has good scientific underpinnings for  sufficiently elevated doses of toxins and exposure to
them~\cite{Kataria15}. The environmental degradation
 arising from  industrialization and the need for intensive  agriculture to feed an increasing
demographic
since World-War II are well recognized. The increasing use of herbicides like glyphosate since
 its introduction (in Sri Lanka circa 2000, and several years earlier to Indian agriculture)
 have been a matter of
public concern. Hence, such considerations would  be the first suspect
in regard to the appearance  of CKDu. However, such a hypothesis has not
found experimental support. A review of the  WHO-sponsored study~\cite{WHO2}, and  other independent
studies~\cite{NanayakkaraS14, Levine16} ultimately lead to the conclusion that agrochemical residues
are below well established thresholds for chronic kidney disease of any kind. 

In the case of glyphosate,
in addition to the usual microbial degradation via the AMPA pathway, glyphosate is removed from the
ecosystem and made into an insoluble, inert chelate~\citep{SmithRay88} of essentially zero bioavailability
due to the action of calcium ions (and similar chelating cations) found in the hard water of the
 endemic regions. The WHO-sponsored study~\citep{WHO2} found no significant levels of glyphosate
 in biopsy studies of CKDu patients in 97\% of the cases, while the 3\% cases with glyphosate
 traces are within the error bars of such field studies.  
 This is one reason why some researchers,
dissatisfied by the conclusion that agrochemical residues seem to be not implicated,
look for ``synergies'' among below-threshold amounts of residues as possibly
 leading to chronic epidemic effects.
Alternatively, they measure the deactivated insoluble chelate in the
sediments and soils and continue to claim that there is persisting ``glyphosate'' in the environment, although
what is found is a very different (ceramic-like) substance, a polymeric
N-(phosphonomethyl) glycinate, and not glyphosate.
 In fact, some researchers had already rushed to name the
disease as an agricultural nephropathy.

On the other hand, contrary to the findings of most authors (e.g.,~\cite{NanayakS19, Levine16})
 Jayalal et al.~\cite{Jayalal19} claim to find
significant above-threshold  amounts of Pb and Cd (but not arsenic) in the diet of
 some 70-80 residents in CKDu-endemic
 villages that they studied. Furthermore, these authors have only studied cadmium or lead, and without
  simultaneously determining the concentration of antagonistic ions like zinc or selenium in the diet.
 Hence, such studies are likely to be totally inadequate for making conclusions about chronic toxicity.
 Furthermore, a common problem in such studies is that  the `total cadmium' determined via mass-spectroscopy
 is not the `bioavailable' cadmium in the food. While other authors have also found
 Cd and other metal toxins  in rice and vegetables (both in the endemic areas, and in other
 areas)~\cite{Premaratna06,WHO2,diyaba-Rice2016,DW2018} their evaluations of typical diets did not lead to a
conclusion about toxicity from the diet. Premaratna's work~\cite{Premaratna06} attempts
to indicate total Cd as well as bioavailable values (about 30\% of total Cd).
  If the individuals in the test sample of Jayalal et al.
 had indeed been affected by cadmium toxicity, in addition to it being reflected in the analytical
 data,  more well established clinical signs of Cd toxicity~\cite{ARLS-review} besides CKDu  should have
 been evident, but no such supporting evidence is indicated in the report by  Jayalal et al.

In Ref.~\cite{DW2015} we argued that large rivers flowing from the agricultural regions of
the tea plantations can bring down macro-nutrients like phosphates found in fertilizers,
but this does {\it not} apply to the micro-components (e.g., cadmium) of fertilizers, as shown in
Sec. 2 of Ref.~\cite{DW2018}. Such rivers flow into all parts of Sri Lanka while CKDu is restricted to
regions with regolith aquifer-fed household wells. 
 
In the case of Ecuadorian nephropathy,
it should not be forgotten that  oil extraction  in Ecuador began in 1972 and became a major pillar
of the economy. Millions of gallons of oil and toxic residues have been discarded directly
 onto the environment causing health and environmental issues \cite{Jochnick94}. Indeed, some 30 billion
 gallons of toxic wastes and crude oil had been discharged into the land and waterways of
 the Ecuadorian Amazon
 by 1993~\cite{Jochnick94}. In contrast, the  use of  agrochemicals is less
 widespread among the poor pastoral farming communities that use  more traditional
 agriculture. Nevertheless,
while agrochemicals like glyphosate have been targeted for ban by zealous activists opposed to
 genetically modified (GM) foods,  little action has been taken against the oil and mining industries. 

Thus, while some proposed causative factors have a good basis in experimental data, most others are
unsubstantiated,  even if strongly pushed by anti-GM activists.   However, Jayasinghe et al. basically
 include  these so-called causative factors 
irrespective of weather there is evidence for them in the traditional sense, e.g., via the
Bradford Hill criteria~\cite{Hill65}, or within  more  modernized criteria 
(e.g.,as discussed in ~\cite{Fedak2015}). The latter are
increasingly  incorporated into mathematical
models and  dynamic toxico-kinetic models. Such mathematical methods enable the researcher to carry out
data integration from a variety of sources or causative agents, and use quantitative measures, e.g., 
for what Hill termed ``strong'' or ``weak'' associations. 

\section{The causative-pathways model for CKDu by Jaysinghe et al.~\cite{Jayasinghe2020}}
Given that the publication of Jayasinghe et al. is a short communication, we may regard 
it  as a declaration of intent for an upcoming more detailed study. Hence we propose to examine here
the limitations of their preliminary formulation as this discussion  may help to pave the way
for a better treatment in future studies.

\begin{figure}[t]
\includegraphics[scale=0.5]{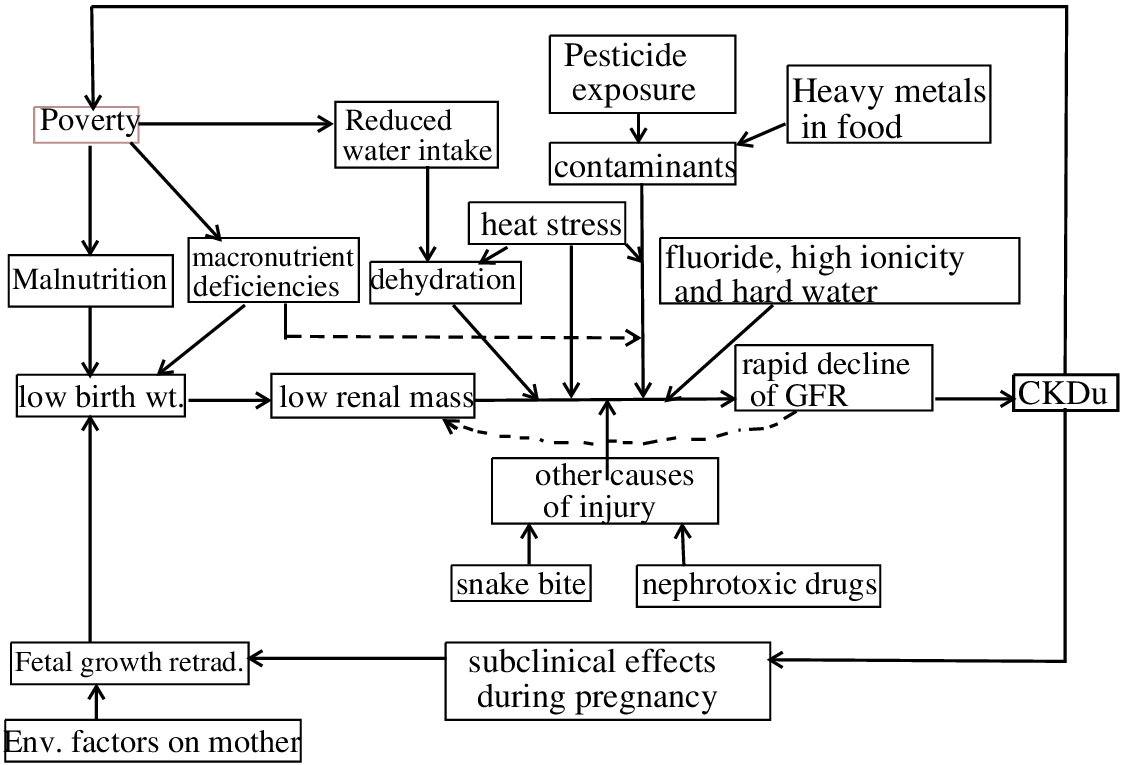}
\caption{The causative pathways of CKDu as presented by Jayasinghe et al.~\cite{Jayasinghe2020}
as a model of a complex system.}
\label{jaya.fig}
\end{figure}
The motivation for the model proposed by Jayasinghe et al. seems to be a somewhat philosophical
 view that CKDu is a disease that cannot be linked to  dominant, easily charcterizable  simple causes.
Thus the authors claim to reject the  ``reductionist'' approach arguing that it fails to  capture certain
emergent modes found in complex systems (CS). It is not clear if Jayasingha et al. believe that ALL the
 causative factors (CFs), even if weak,  indicated in their organigram (see Fig.~\ref{jaya.fig})
 model of the organism and the environment as a  complex interacting system (MCS) are considered
 to be {\it necessary and sufficient} to cause CKDu. 
No clarity is afforded as to whether certain CFs, or all displayed
act together  and become synergistic. Furthermore, it is surprising to find that no antagonistic factors are
included in their MCS. Thus, the  presence of  concentrations of
Zn$^{++}$ ions that significantly exceed
the concentration of Cd$^{++}$ ions possibly  present in the dietary intake is known to largely
nullify the usual nephrotoxicity of cadmium~\cite{ARLS-review,Chaney12,DW2018}. This antagonistic factor
 is not included, and constitute one example of a serious lacuna found in the MCS of Jayasinghe et al.

In chapter 9, section 3 of the Ref.~\cite{apvmm2013} we have discussed how standard methods of 
scientific investigation recover  emergent modes quite easily. Very often, the emergent mode
can be identified without a complicated systems analysis as many feed-back loops can be easily
 identified. In the
case of chronic kidney disease (CKD), a trivial emergent mode is seen in the feedback loop connecting the
decreasing renal mass and  increasing glomeruler and tubular damage. This is embodied in the
``hockey-stick shaped''  emergent characteristic of the plots  displayed  in the toxico-kinetics
 of chronic kidney disease. Thus the toxic effects are very weak and almost linear as a function of
 time in the
early stages of the disease, but sharply rise beyond stage two of CKD and most probably in CKDu as well.
This is due to feed-back effects coming into play.
While one may identify this particular feed-back loop in Fig.~\ref{jaya.fig} which connect the box labeled
``low-renal mass'' (LRM) to the box labeled ``rapid decline in GFR'' (RDGFR) and back to ``low-renal mass'',
 it does not provide a means to construct the corresponding toxico-kinetic equations or determine
 the emergent modes.

Two other trivial feed back loops are found in the MCS figure of Jayasinghe et al.
 One of them is the upper loop
going from low birth weight (LBW) to CKDu and then, through ``poverty'' and ``malnutrition''
 to LBW of a new born.
So the time scale of this loop is of the order of a trans-generational time $\tau_g$ . A similar
lower  loop goes through the box labeled ``subclinical effects during pregnancy''.  This also has a
 trans-generational timescale. However both these loops, as well as the renal-mass loop
 suggest that within time scales $\tau < \tau_g$ the mother is also very weakened, and vulnerable to CKDu
or other diseases,
and hence may not live long enough to give  progeny. Hence these two loops become
unimportant. Further more, the very structure of the complex system proposed by Jayasinghe et al. leads us
to the view  that women are more likely to contract  CKDu. In fact, most field studies~\citep{Ranasinghe19}
 suggest that men are
 more likely to contract CKDu than women although other views exist~\citep{Kafle19}.

We may also note that no other emergent modes except the ones enumerated here are likely to  arise in the
model of Jayasinghe et al. This is discussed further in sec.~\ref{emerg.sec}.
 The authors  themselves have not indicated any
 emergent modes while hinting that emergent modes
may arise. Presumably, the causal factors associates with the five arrows that direct to the loop connecting
 LRM to RDGFR cause the rapid decline in renal function. But how this happens, and if they are even
relevant, have to be assessed elsewhere. In act, the only experimentally established synergies for
 Sri Lankan CKDu 
are those established by Wasana et al.~\cite{SynergyBandara2017} in research carried out by the team
of  Banadara et al.

For more  complex situations, a  particularly useful approach
is afforded  by artificial intelligence (AI) and  neural network-type models. They
 enable one to ``train'' a neural network and  improve on its predictive quality as more and more
 new data are fed into it. A short introduction to such neural networks in a general context  may be
found in Chapter 2, Sec.~2.5 of Ref.~\cite{apvmm2013}. The type of field data needed for such studies
 is not currently available for CKDu.

Another short-coming of the Jayasinghe model is that it does not take local effects into account.
Its components can can apply to any part of the country, 
{\it except} for
the box labeled ``fluoride, high ionicty and hard water''. That only applies regolith-aquifer fed  areas
 in the country. CKDu has been found in those areas, previously jungle, after they were settled under
 the Mahaweli irrigation scheme in the 1970s. Furthermore, no CKDu is found in all the areas
 where that box is not applicable, even though all other boxes, e.g., poverty, malnutriton,  contamination,
low birth weight etc., are  all found in other parts of  the country where manual workers
labour under the tropical sun. Interestingly,  animals that consume agricultural water (in paddy fields, canals)
are subject to all the causative factors in the Jayasinghe-Zhu model, except high-ionicity well-water
containing F$^-$ and Mg$^{++}$, 
and they do not contract CKDu.

\paragraph{"Multifactorial" models}-- Given that various investigators of CKDu have proposed many possible
causes, some authors have suggested that the etiology of the disease is `multifactorial'~\cite{Wimalawansa16}.
 Such proposals are less ambitious than the attempt to set up a dynamical multifactorial
 model~\cite{Jayasinghe2020}.  However, the mere fact that various authors have proposed various
 (different) causes  does not by itself  establish a maltifactorial etiology for a disease. The model
given by Wimalawansa is currently lacking in any detail and  use  categories like ``Environment'',
``unidentified toxins'',  etc., and hence it is not too different from saying that the etiology is unknown.

In our view, the evidence available so far suggest that CKDu is caused by a single aeteological factor,
namely, the simultaneous occurrence of fluoride and magnesium ions (present as a component of hardness) in
 drinking water drawn from regolith aquifers.

In effect, the MCF presented by Jayasinghe et al. can be regarded as an improvement over previously
presented ``multifactorial' models.  While there are
many factors that cause CKD or CKDu in different situations, and hence require different strategies
 for their prevention, the claim of multifactorial origin should strictly mean that we know,
 and can identify,  several  factors  that definitely contribute to the origin of the diseases 
in that those causative factors are {\it necessary and sufficient} for causing CKDu.
 When such information is not available, including a variety of
factors in an  organigram or changing the name to `multifactorial origin', or to
 ``agricultural nephropathy'' 
 can be quite misleading.  That the complex-system model can be easily misused
 is best explained using a counter-example.

\section{A counter-example based on the Jayasinghe et al.  MCS}
\begin{figure}[t]
\includegraphics[scale=0.5]{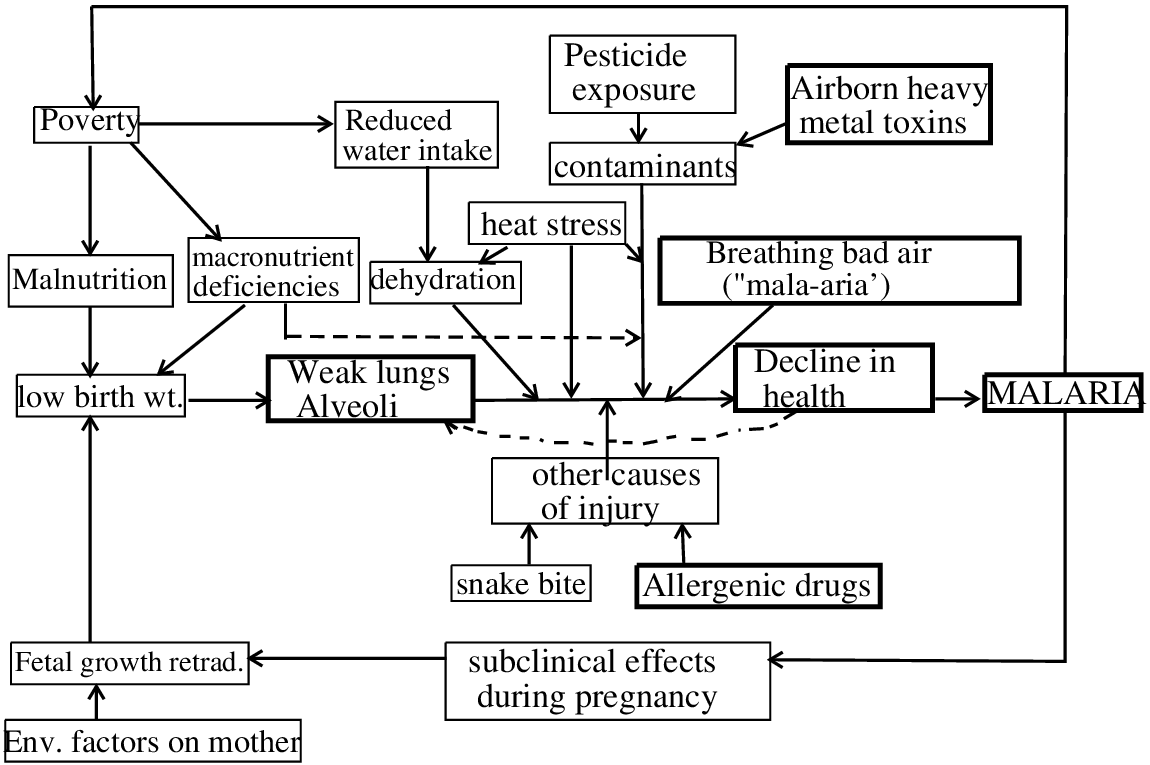}
\caption{A causative-pathway scenario for Malaria following the model
 of Jayasinghe et al.~\cite{Jayasinghe2020}
with malaria as an ``emergent property'' of the complex system, showing the propensity for
 misleading conclusions.}
\label{malariaMCS.fig}
\end{figure}

In this section we reuse the causative pathway model for CKDu given by Jayasinghe et al.,
 and slightly modify it
to apply to a hypothetical attempt to understand the etiology of malaria (or lung cancer)
 imagining   a time where their
origins in the mosquito vector, (or tobacco smoking) were not understood.
 The modified flow diagram is given
in Fig.~\ref{malariaMCS.fig} for malaria. Those boxes which have been modified from the
 CKDu model of Jayasinghe et al.  are marked with thicker lines. A similar modification
 can be done appropriately for lung cancer to propose that an emergent multifactorial
 mechanism, and not  tobacco smoking, is the cause of lung cancer!   

In effect, using such a model we can  argue that malaria is an emergent property of the environmental
factors associated with various types of air pollution and ``bad air''. Prior to the recognition that 
malaria was spread (or `caused' in popular parlance) by mosquitoes, the Italian name ``mala-aria''
 revealed the commonly accepted belief about the etiology of the disease. This counter example illustrates
 the danger of
 using such `multi-factorial' models as ``explanations''  inclusive of `lots of small causes working together'
to  `explain' the etiology of a disease.

\paragraph{Conditions for several causative factors to add together.}-- Causative factors can add together
and act synergistically when suitable microscopic mechanisms exist. Additionally, one process may be able to
provide energy to push another process forward, and augment the process. However, if the number of weak processes
is large, then a form of the central-limit theorem comes into play, and instead of synergistic action, we
end up with processes that merely contribute to `white noise' while only the effect of the  dominant factor
will appear, distributed as a Gaussian. A more formal discussion of
emergent modes  is given in the following section.

\section{Application of a formal theory of emergent modes to the current model.}
\label{emerg.sec}
In the following we apply standard coupled-mode theory to discuss emergent modes
in the problem of the response of an organism to
environmental causes, within a complex system containing at least one feedback loop.
As already discussed, there is just a single feed-back loop of any relevance
 in Fig.~\ref{jaya.fig}. This is the
loop connecting ``low-renal mass'' to ``rapid decline of GFR'' and back. There are five processes that
are depicted in Fig. 1 as  contributing to it, via five arrows. Identifying each arrow by the box
 directly connected
to it, these are (1) dehydration, (2) heat stress, (3) other causes of injury, (4) contaminants, and 
(5) fluoride and ionicity. Of these, in our opinion, the first four have negligible weight
 and are irrelevant. We assume that the variables have been orthogonalized in the standard  way by an
eigenvector analysis, or factor-group analysis. For simplicity of discussion, let us assume that
 the orthogonalization
step did not mix the causative agents strongly, so that the original choice of causative agents
are already effectively independent causative factors. 
Let us assume that they are all included with weights $w_i, i=1,5$. Their inputs into the feedback
 loop are described by the vectors $\vec{X}_i(t)$ at the time instant $t$. That is, each causative factor is
associated with its input vector $\vec{X}_i$.  The effect of each causative factor
 on a chosen kidney function characteristic (e.g., GFR)  is denoted  by $F(t)$. 

 $F(t)$ changes by an amount $\Delta_i F(t)$ under the $i$-th  causative factor. Then,
if there is no feed-back loop action, we can formally  write the
 response of the renal
 system to  the $i$-the environmental factor by the relation
\begin{equation}
\label{zeroth.eqn}  
\Delta_i F(t)=\xi^0_i(t)\vec{X}_i(t)
\end{equation}
Here $\xi^0_i$ is the ``zeroth order''  response of the renal system to the $i$-th environmental
 causative factor. As we are dealing with ``chronic'' effects, i.e.,  effects that act  over a
length of time, unlike acute toxicity, these effects are weak but cumulative. Hence the use
of a linear-response form is probably well justified. 
As no feedback effects or interactions between causative agents are included at this stage,
 we call $\xi^0_i$  the zeroth-order
response. This can be determined experimentally by monitoring the change in renal function under the action
of the given environmental factor. The above equation is the effect of just the $i$-th causative agent.
Thus the total effect of all five causative factors acting on the renal system with no feed back effects, or
synergistic or antagonistic effects can be written as
\begin{equation}
\Delta F(t)=\Sigma_i w_i\xi^0_1(t)\vec{X}_i(t)
\end{equation}
To proceed further, we assume that each causative agent acts on the renal system
via an interaction factor $V_i$  associated with the time scales $\tau_i$. Furthermore, we  transform 
the variables $(\vec{X}_i, t)$ to work in the Fourier  space of  $(\vec{K}_i,\omega)$ where $\omega$
is a frequency. Then the time scales $\tau_i$ are replaced by their corresponding frequencies $\omega_i$.
Then the total response $\xi_T(\omega)$  of the renal-GFR feed-back loop to all five environmental
agents can be written as:
\begin{equation}
\xi_T(\omega)=\Sigma_i w_i\frac{\xi^0_i(\vec{K}_i,\omega)}{1-V_i(\vec{K}_i,\omega)\xi^0_i(\vec{K}_i,\omega)}
\end{equation}    
In the approximation where the effect of each environmental factor is characterized {\it only} by the
time-scale $\tau_i$ that it sets on the life of the renal system, the above equation can be simplified
and written in the form:
\begin{equation}
\label{sum.eq}
\xi_T(\omega)=\Sigma_i w_i\omega_i^2/(\omega^2-\omega_i^2)
\end{equation}
Note that each term has a pole at $\omega=\omega_i$ so that the lifetime of the renal system under the
$i$-the environmental insult is $\tau_i$.
At this level of analysis, the environmental factors are acting additively  on the renal-mass-GFR loop
without any synergistic or antagonistic effects  included. To include such effects, for simplicity, let us consider 
just two  environmental agents to become synergistic, say the pair  $i=1$ and 2, and let their interaction
 be denoted by $V_{12}(\vec{K},\omega)$. If there are no interactions between a given pair, the corresponding
 $V_{ij}$ is zero. These can be regarded as  toxico-kinetic terms that have to be determined
from experiment or from physico-chemical energy considerations ~\cite{DW2017}, or possibly from
a fundamental model of biochemical processes. Here we simply assume it to be available  data characterizing
the complex system. Then
 the synergistic or
antagonistic effect changes the first two terms of the equation~\ref{sum.eq} and gives the form
\begin{equation}
\xi_T(\omega)= \frac{w_1\xi^0_1(\omega)+w_2\xi^0_2(\omega)}
{\left[\{1-V_1\xi^0_1\}\{1-V_2\xi^0_2\}-\xi^0_1|V_{12}|^2\xi^0_2\right]} + i>2\; \mbox{terms}.
\end{equation}
The new denominator no longer has poles at $\omega_1$ or $\omega_2$. Instead, new poles  are given by the
roots of the equation
\begin{equation}
\left[\{1-V_1\xi^0_1\}\{1-V_2\xi^0_2\}-\xi^0_1|V_{12}|^2\xi^0_2\right]=0.
\end{equation} 
These roots define  the new, or ``emergent'' modes resulting from the synergistic
 or antagonistic interactions, and give new time scales $\tilde{\tau}_1$ and  $\tilde{\tau}_2$.
If the interaction is antagonistic, the time scales $\tau_1,\tau_2$ corresponding to the
 original poles $\omega_1,\omega_2$
get lengthened to such an extent that renal damage will not occur within the patient's life time. On the
other hand, if the causative factors act synergistically, then the lifetimes $\tau_1,\tau_2$ will
 be shortened
significantly and will also shorten the lifetime of the patient.

\subsection{Can many below-threshold effects contribute to become a significant emergent cause?}
The zeroth-order time scales $\tau_i$  or their corresponding frequency values $\omega_i$ of below-threshold 
are such that their effect would be felt only if a human being lived to a thousand years or so. So, even
if a hundred of them joined together to give new renormalized time scales $\tilde{\tau_i}$, they will still
be too far out to have any effect within the lifetime of a patient. In fact, the logic behind setting
thresholds of chronic toxicity is precisely that no toxicity was observed and not expected to be observed
under long term monitoring. It is always usually the case that besides the many below-threshold effects,
there are usually a few dominant effects that carry a high statistical weight based on the available 
evidence.

In Fig.1 the statistical weights $w_i$ seem to be negligible except for the fifth
case, imposing the reductionist picture by the weight of evidence. 
That is, the action of consuming fluoride and magnesium ions found in wells supplied by regolith aquifers
prevalent in the endemic regions is by far the most significant causative environmental factor for CKDu.

\section{conclusion}
The construction of models of complex systems, as well as the mechanisms of origin of emergent modes  is
 well understood and requires much more stringent and accurate data than are needed for models that are derived
from a reductionist approach where the number of variables is strictly controlled by designing suitable
experiments. Complex systems (see chapter 2, Ref.~\cite{apvmm2013}) can be built up with confidence only when
experimental information is available on a large  number of reductions of the more complete  system. This is
in fact a major problem in environmental studies.
These limitations  of modeling complex systems make it 
extremely easy to obtain false conclusions from `complex-system' models. 
  
\section{Funding} 
This research did not receive any specific grants or financial support
in any form from funding agencies in the public, commercial, or not-for-profit sectors.

\section{Declaration of competing interests}
The author has  no known competing financial, commercial or
personal interest in  the work reported here.

\bibliography{mybibfile}

\end{document}